\documentclass[12pt]{iopart}
\usepackage{iopams}
\usepackage{epsfig}
\usepackage{graphicx}% Include figure files
\usepackage{dcolumn}% Align table columns on decimal point
\usepackage{bm}% bold math
\def\be{\begin{eqnarray}}
\def\ee{\end{eqnarray}}
\def\ba{\begin{array}}
\def\ea{\end{array}}

\begin{document}
%\begin{frontmatter}
\title[IQHE topological insulator]{Interaction reconstructed integer quantized Hall effect topological insulator}
\author{A. Yildiz $^1$, D. Eksi $^2$ and A. Siddiki $^2$}
\address{$^1$Department of Physics, Faculty of Sciences, Dokuz Eylul University, Tinaztepe Campus, 35160-Buca, Izmir, Turkey}
\address{$^2$Department of Physics, Faculty of Sciences, Istanbul University, 34134-Vezneciler, Istanbul, Turkey}
\ead{afifsiddiki@gmail.com}
\begin{abstract}
We discuss the role of direct Coulomb interaction on the bulk insulator of the integer quantized Hall effect that bridges the topological insulators and the conductance quantization. We investigate the magneto-transport properties of a two-dimensional electron system in the bulk, numerically, utilizing the self-consistent Thomas-Fermi-Poisson screening theory. Topologically distorted Hall bars with and without potential fluctuations are considered that comprises two identical inner contacts. Although these contacts change the topology, we show that quantized Hall effect survives due to redistributed incompressible strips on account of interactions. It is shown phenomenologically that the impedance between these contacts can be obtained by a minimal transport model. An important prediction of our self-consistent approach is a finite impedance between the inner contacts even in the plateau regime, where the maximum of the impedance decreases with increasing temperature.
\end{abstract}

\maketitle
%****************************************************
%\begin{keyword}
% keywords here, in the form: keyword \sep keyword
%quantum Hall effect\sep topology
% PACS codes here, in the form: \PACS code \sep code

%\end{keyword}
%*****************************************************
%\end{frontmatter}
\section{Introduction}
Since the discovery of the integer quantized Hall effect (IQHE)~\cite{vKlitzing80:494} in a two dimensional electron system (2DES), its topological properties are
investigated intensively~\cite{Thouless82:405,Niu85:3372,Hasan:10:3045}. The guiding gauge invariance
argument of Laughlin is fundamental to the phenomena, which focusses on a cylinder considering a strong, time-dependent
magnetic field $B$ threaded to its surface in the normal direction~\cite{Laughlin81}. Each time the field is increased by one
magnetic flux quantum ($\Phi_{0}=e/h$, $e$ the elementary charge and $h$ being the Planck constant), an electron is argued to be
transferred adiabatically from one side of the cylinder to the other side, protecting the geometrical phase.
The result is a flow of electrical current proportional to the
electromotive driving force, with a precisely quantized
proportionality coefficient- the Hall conductance. The topological aspects of the IQHE assuming periodic boundary conditions was first discussed by Thouless {\emph{et al}}, where Hall conductance is re-introduced as the topological invariant of the 2+1 D system~\cite{Thouless82:405,Niu85:3372}. In these highly appreciated
works, the topological character of the 2DES is described in
terms of the incompressible bulk state and an expression is obtained via Kubo formalism for the Hall
conductance, which is in turn determined by the Chern number~\cite{Hasan:10:3045}. The topological arguments rely on the fact that
the Hall conductivity is an integer multiple of $e^2/h$, if the Fermi energy $E_F$ is in between
two Landau levels. In this situation the 2DES is incompressible in the bulk, namely there are no available states at $E_F$. Within the non-interacting picture, charge transport along the edges is described by dissipationless 1D edge channels~\cite{Halperin82:2185,Buettiker86:1761}, which completes the analogy between the topological insulators and IQHE. It is very important to emphasize that all the above descriptions of the IQHE are in the \emph{momentum} space, $k$. To be explicit, one assumes periodicity in $k-$space, perform calculations related with geometrical phase in this space and then due to the symmetry arguments maps momentum space directly to real space. For sure, such a mapping can be justified only if the electronic system is periodic both in momentum and in real space. For instance if one performs calculations on a closed cylinder, with a radial $B$ field. However, when considering a real size sample with finite length and perpendicular field the symmetry argument is prone to fail. In addition, disorder emanating from remote donors provide localized states which result in finite width quantized Hall plateaus in certain $B$ field intervals. Hence, the incompressibility of the bulk throughout this interval is preserved and is essential to observe IQHE. Namely, without disorder the plateaus would shrink to a single $B$ value for an unbounded 2DES, where the Landau level filling fraction $\nu$ is an integer, which is defined by $\nu=N_{\rm{el}}/N_{\Phi_0}$~\cite{Kramer03:172}. Here, $N_{\rm{el}}$ is the number of electrons
in a certain area $A$ and $N_{\Phi_0}$ is the number of flux quanta in $A$. The incontestable concept of insulator bulk state has been challenged by the inclusion of interactions starting from early 1990's advocated by Chang~\cite{Chang:90}. Later, Chklovskii and co-workers analytically showed that the direct Coulomb interaction results in a metal-like (compressible) bulk, whereas the insulating states can also be at the edges suppressing back (or forward) scattering within the plateau interval~\cite{Chklovskii92:4026}. In the last decade, the bulk and edge electrostatic properties of the 2DES is investigated with the improvement of local probe techniques~\cite{Ahlswede02:165,Dahlem:10,Suddard:10}. Measurements show that the spatial distribution of the theoretically predicted edge incompressible strips and the experimentally observed poor screening regions coincide nearly perfect. However, with a crucial qualitative difference: the edge strips become transparent to Hall field whenever the widths of these incompressible strips become comparable with quantum mechanical or thermodynamical length scales. This discrepancy is removed when the electrostatic problem is solved also taking into account quantum mechanical quantities, like the finite widths of the wave function, together with the effects of finite temperature and level broadening~\cite{Guven03:115327,siddiki2004}.
\begin{figure}[t!]{\centering
\includegraphics[width=.26\columnwidth]{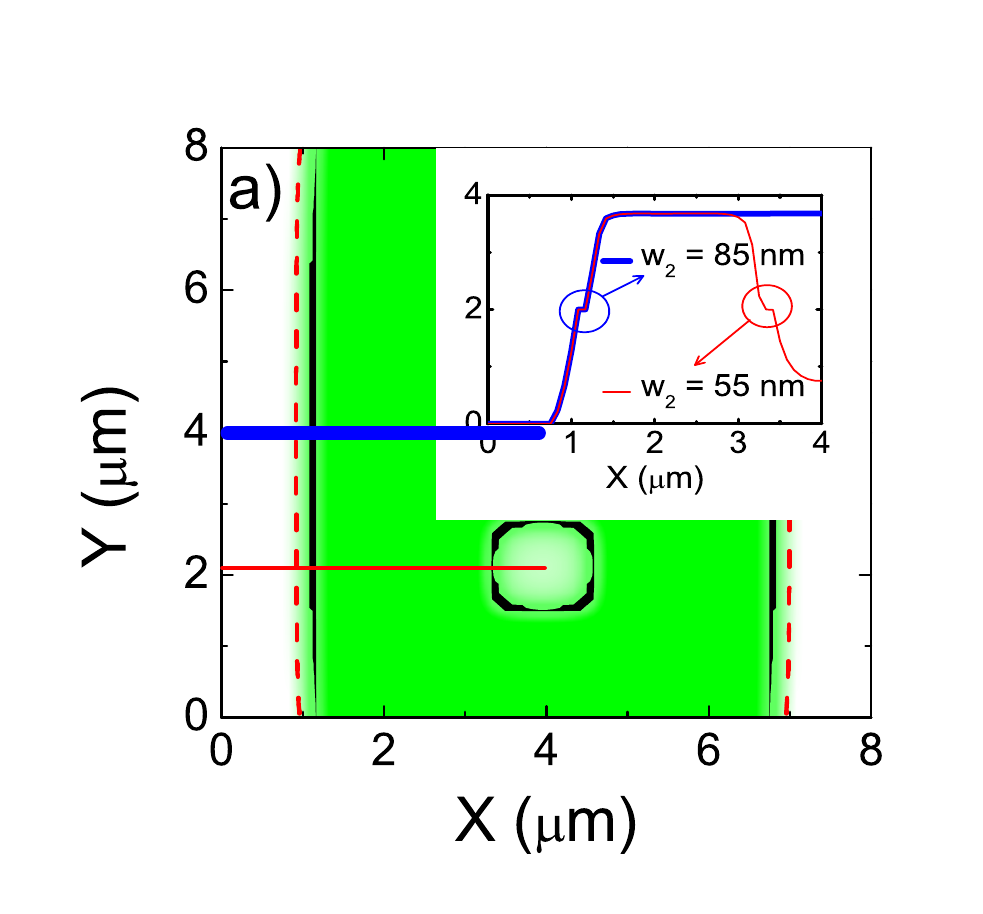}\hspace{-.6cm}
\includegraphics[width=.26\columnwidth]{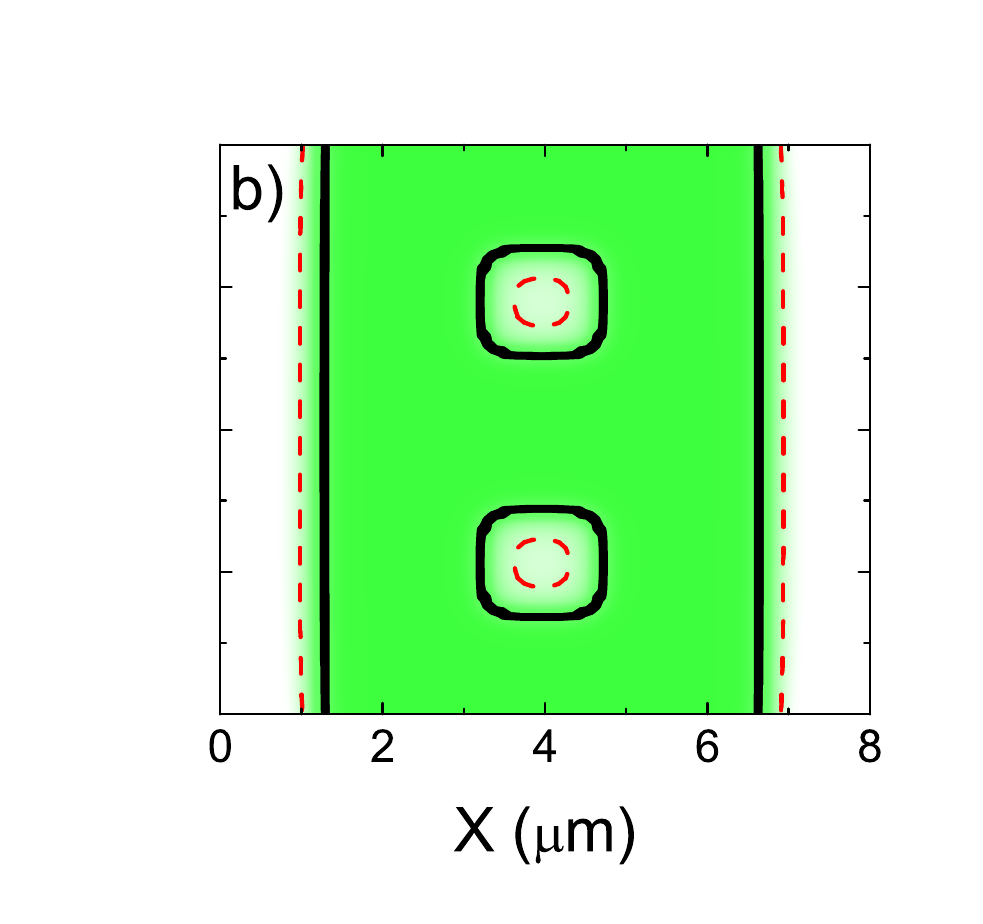}\hspace{-.6cm}
\includegraphics[width=.26\columnwidth]{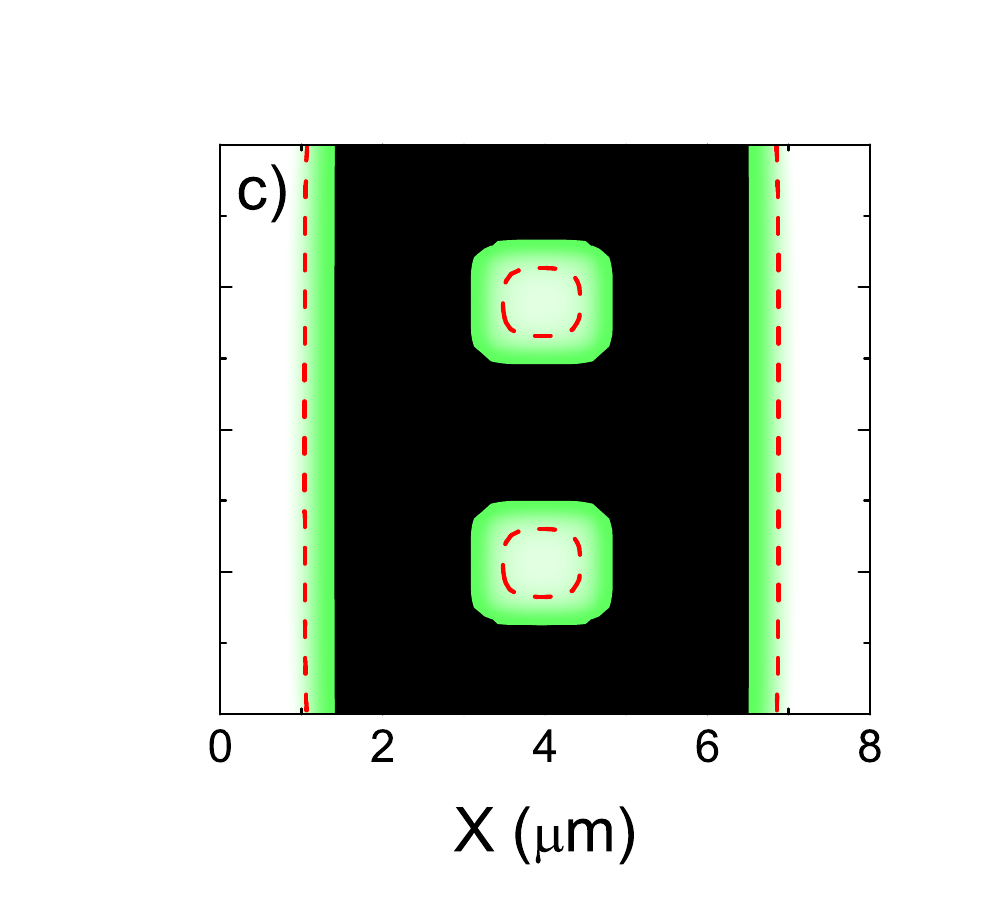}\hspace{-.6cm}
\includegraphics[width=.28\columnwidth]{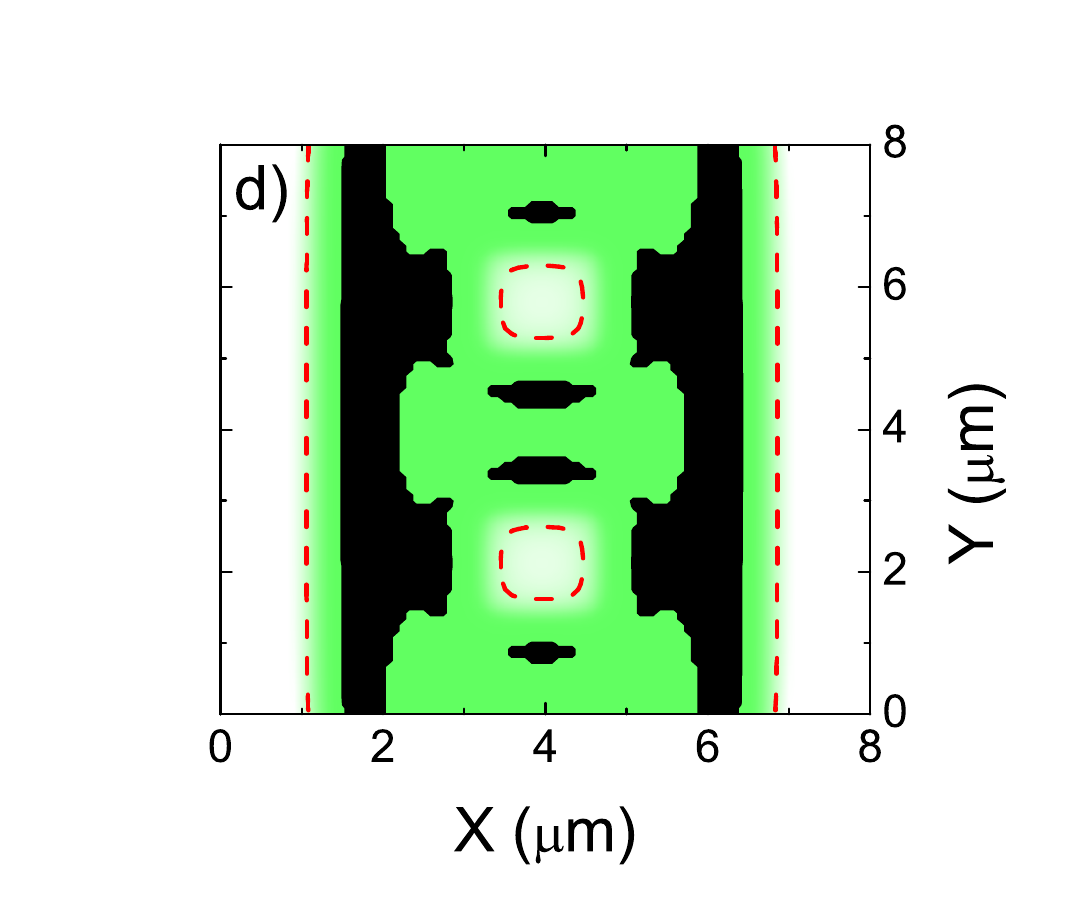}\hspace{-.6cm}
\vspace{-0.5cm}\caption{\label{fig:hall_bar} Filling factor distribution in the Hall bar for characteristic $\Omega_c=\hbar\omega_c/E_F^0$ values, where $E_F^0$ is the Fermi energy in the bulk calculated for a typical electron density of $3\times10^{11}$ cm$^{-2}$. a) $\Omega_c=0.81$, b) $\Omega_c=1.31$, c) $\Omega_c=1.45$ and d) $\Omega_c=1.5$, calculated at sufficiently low temperatures, $\tau=kT/E_F^0=0.016$. The inset depicts the spatial variation of the density along the vertical lines.}}
\end{figure}

The purpose of this paper is to discuss the role of interactions on the bulk insulator of the IQHE by solving the Schr\"odinger-Poisson equation in 2D coordinate-space, self-consistently. By implementing two inner contacts, we investigate the magneto-transport properties of the bulk of a Hall bar by calculating the spatial redistribution of the insulating states which can reside both at the edges or at the bulk. We find that, the inner contacts leave the QHE unaffected although the topology of the system is remarkably changed from genus 0 to 2. Next, we explicitly include disorder to our self-consistent screening calculations by an additional impurity potential resulting from ionized donors. It is observed that, the edge and bulk incompressible regions merge due to potential fluctuations resembling the topological bulk theory of the IQHE. We finalize our investigation by calculating the impedance between the inner contacts by a minimal transport model, which predicts a measurable potential difference between them even within the plateau interval. This prediction imposes that the bulk of a Hall bar is \emph{not} an insulator throughout the plateau interval, hence challenges the analogy between the topological insulators and the IQHE. Our predictions can be easily tested in experiments together with the temperature dependency.

\section{The model} Assuming that the electrostatic quantities vary slowly on the
quantum mechanical scales such as magnetic length ($\ell=\sqrt{\hbar/eB}$), we employ the well appreciated Thomas-Fermi
approximation to obtain the electron density and potential distributions~\cite{Guven03:115327,Lier94:7757,Sefa08:prb}. Then the center coordinate $X$ ($=-\ell^2k_y$, where $k_y$ is the quasi-continuous momentum in $y-$) dependent eigenvalues can be calculated in the lowest order
perturbation as $E_{n}(X)\approx E_{n} + V(X), E_{n}=\hbar\omega_{c}(n+1/2),$ where $\omega_{c}=eB/m^*$ is the cyclotron frequency, $V(X)$ is the potential (energy) and $n$ is the Landau index. We calculate the electron density $n_{\rm el}(x,y)$ and the total potential energy $V_{\rm tot}(x,y)$ from the following self-consistent equations:
\begin{equation}\label{nelec}
n_{\rm el}(x,y)=\int
dE\frac{D(E)}{e^{[E+V_{\rm tot}(x,y)-\mu_{\mathrm{elch}}^{\star}]/k_{B}T}+1},
\end{equation}
and
\begin{equation}\label{vpot}
V_{\rm tot}(x,y)=V_{\rm ext}(x,y)+V_{\rm int}(x,y),
\end{equation}
where $V_{\rm ext}(x,y)$ is the external and $V_{\rm int}(x,y)$ is
the Hartree interaction potential (energy) between the electrons in the system is given by
\begin{equation}
V_{\rm int}(x,y)=\frac{2e^2}{\overline{\kappa}}\int
K(x,y,x^{\prime},y^{\prime})
n_{\rm el}(x^{\prime},y^{\prime})dx^{\prime}dy^{\prime}.
\end{equation}
In above equations, $D(E)$ is the density of states, $\mu_{\mathrm{elch}}^{\star}$ is the electrochemical potential being constant
in the absence of an external current, $\overline{\kappa}$ ($\sim 12.4$, for GaAs) is an average
dielectric constant and $K(x,y,x^{\prime},y^{\prime})$ is the solution of the Poisson equation considering periodic boundary conditions~\cite{Morse-Feshbach53:1240}.
Note that the electrochemical potential in equilibrium is defined as $\mu_{\mathrm{elch}}^{\star}=\mu-|e|\phi(x,y)$, where $\mu$ is the chemical potential and $\phi(x,y)$ is the electrostatic potential.
In addition, the chemical potential is determined by the statistical description assuming
a grand canonical ensemble which is in contact with a reservoir. Starting from
$T=0$ and $B=0$ solutions, we obtain $n_{\rm el}(x,y)$ iteratively in
thermal equilibrium, keeping the donor density $n_{0}$ distribution fixed and average electron density constant, i.e. $\mu$ is constant and position independent. Since our results are independent of the particular nature of the single particle gap, we assume $g_{s}=2$.
\begin{figure}[t!]{\centering
\includegraphics[width=.26\columnwidth]{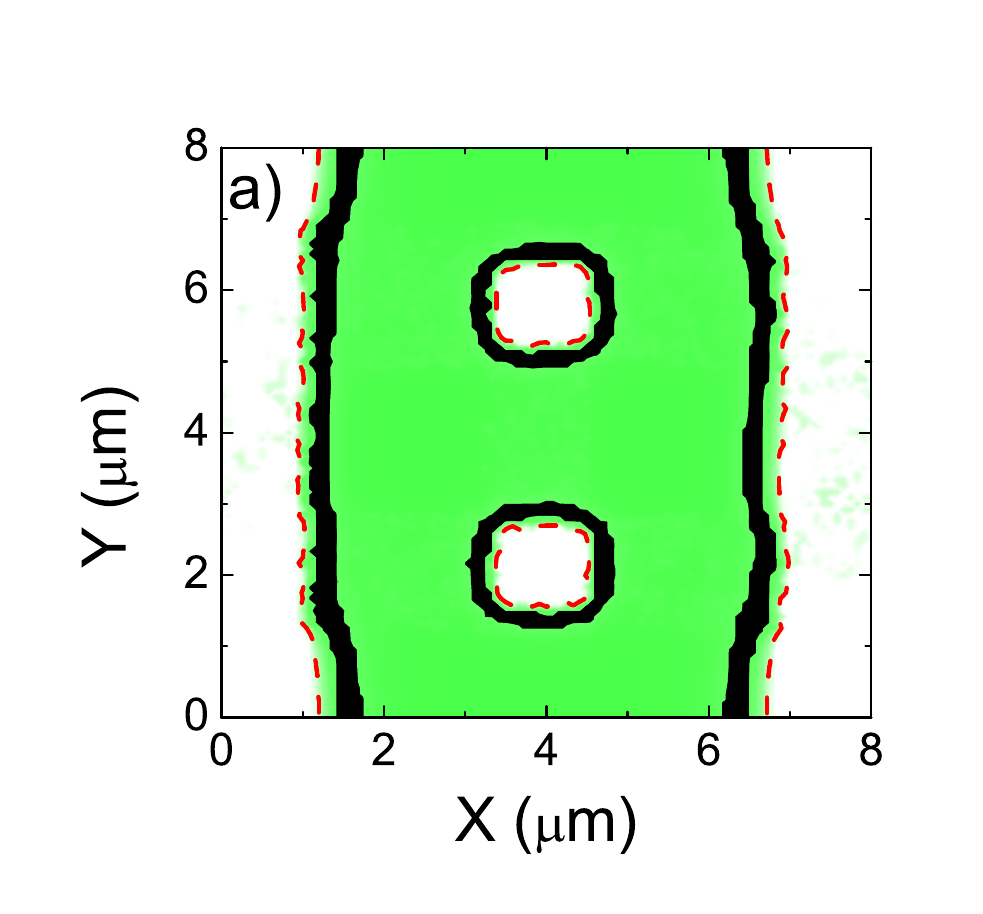}\hspace{-.8cm}
\includegraphics[width=.26\columnwidth]{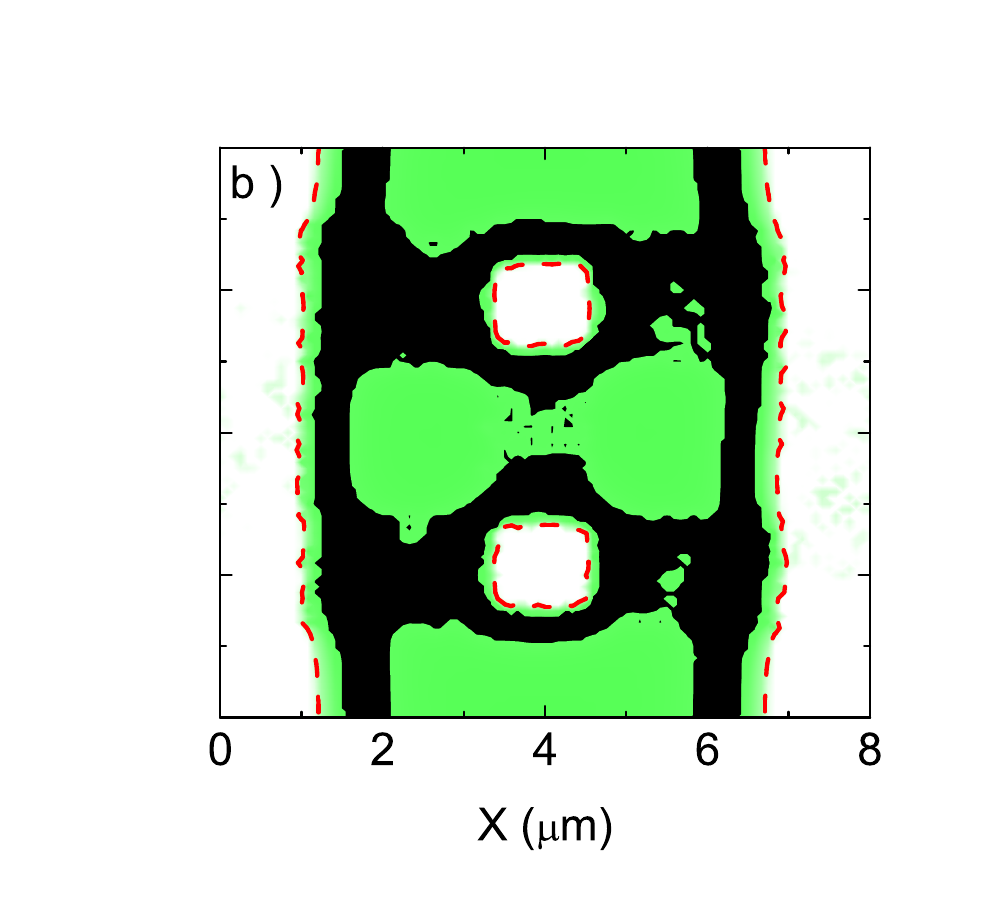}\hspace{-.6cm}
\includegraphics[width=.26\columnwidth]{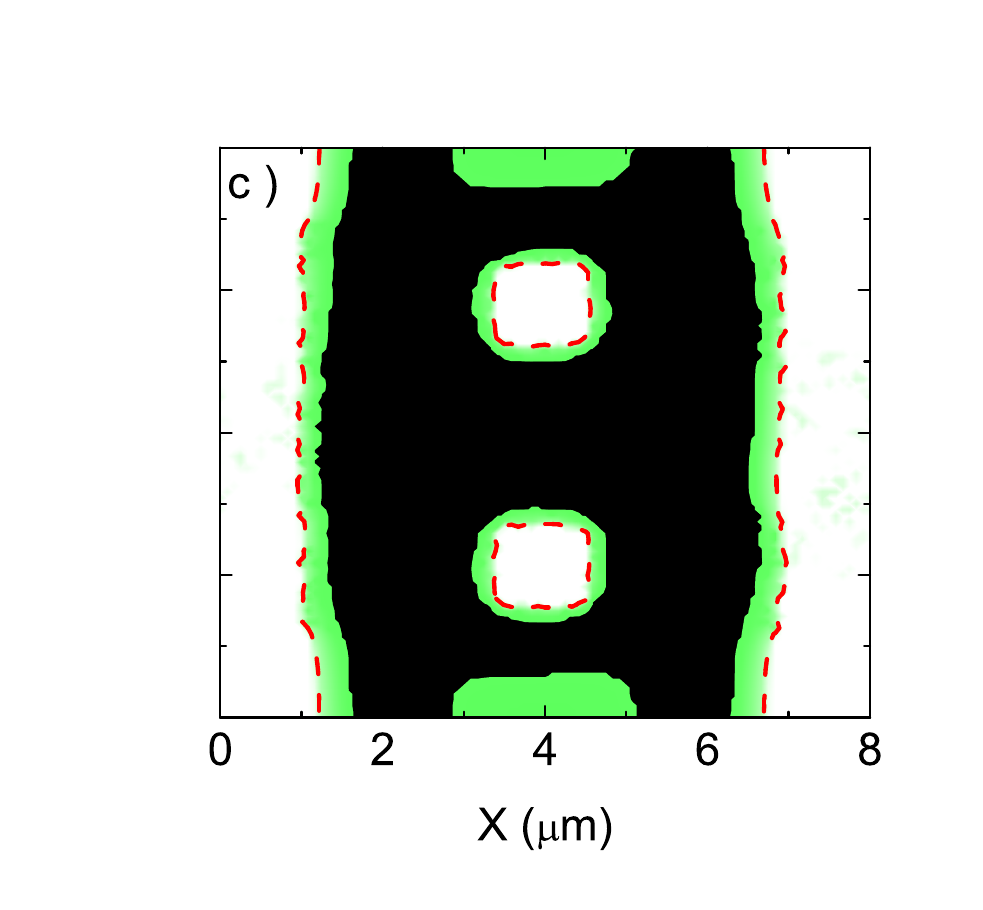}\hspace{-.6cm}
\includegraphics[width=.28\columnwidth]{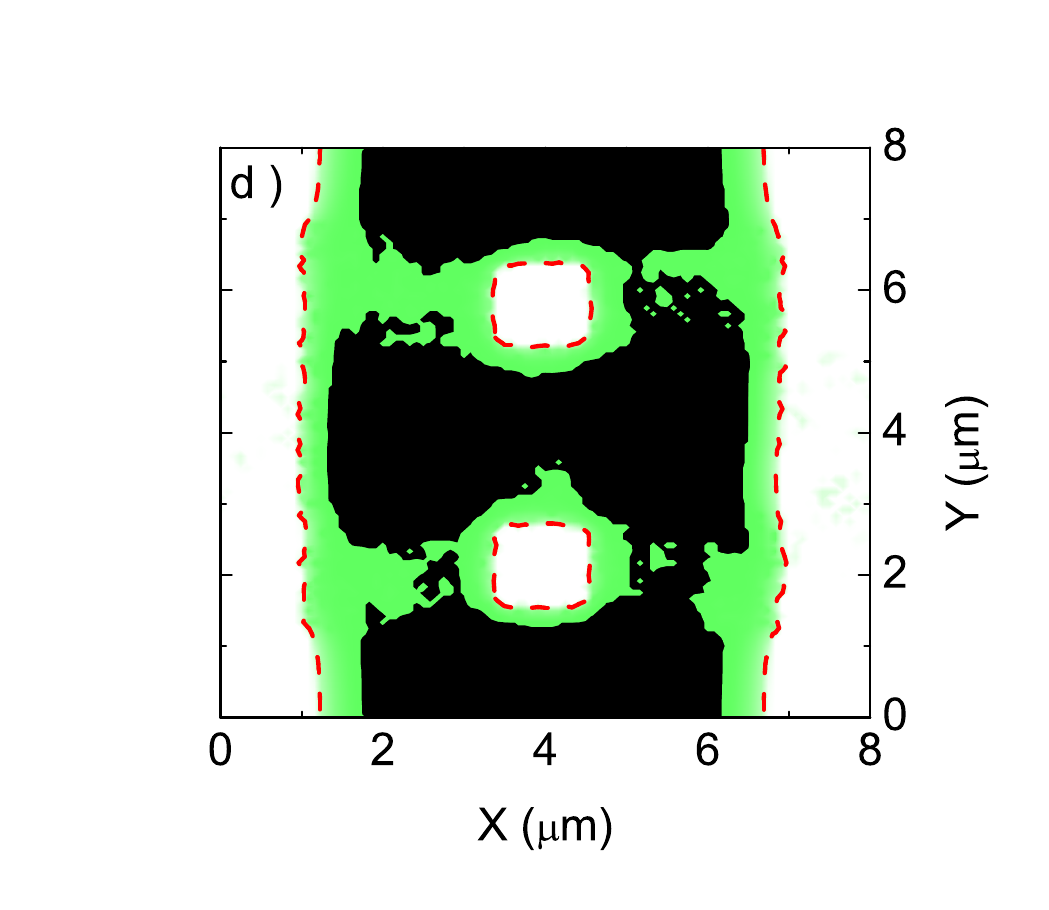}\hspace{-.4cm}
\vspace{-0.5cm}\caption{\label{hallbar_disorder} Same as Fig.~\ref{fig:hall_bar}, however, with additional potential fluctuations emanating from ionized donors. The electron density is therefore slightly ($\sim \% 20$) increased, resulting in field strengths of a) $\Omega_c=1.36$, b) $\Omega_c=1.46$, c) $\Omega_c=1.49$ and d) $\Omega_c=1.52$ calculated at default temperature.}}
\end{figure}

\section{Results and Discussion} To discuss the role of interactions on the bulk insulator of the quantized Hall effect, we will investigate the electron distribution of a clean Hall bar (i.e. without potential fluctuations resulting from ionized donors) and a disordered Hall bar both comprising two inner contacts. In both cases, we will focus on the formation of the incompressible regions and investigate whether if they reside at the bulk or at the edge. In the presence of a strong disorder potential, we will show that bulk and edge incompressible strips merge.

\subsection{The clean Hall bar} We consider a $d=7$ $\rm \mu m$ wide Hall bar without potential fluctuations induced by disorder on which two identical square contacts (of size $\sim1$ $\rm\mu m^{2}$) are defined in the interior. The properties of the heterostructure and the details of 3D self-consistent calculations are given elsewhere~\cite{deniz:contact}. In  Fig.~\ref{fig:hall_bar}, we show the electron density distribution (or equivalently local filling factors, $\nu(x,y)=2\pi\ell^2n_{\rm el}(x,y)$) as a function of spatial coordinates, where the color gradient depicts compressible and black regions correspond to incompressible regions with constant electron density. The electron density effectively vanishes beneath the inner contacts and at the edges (white areas), which is bordered by broken (red) lines. Along the incompressible regions the electrostatic potential varies due to poor screening and is approximately flat when the region is compressible, i.e. $\nu(x,y)$ varies. Since, the formation of incompressible regions is shown to be sensitive to temperature
and the strength of the magnetic field, we will consider situations where the electron temperature is sufficiently low so that the incompressible regions are well developed. Namely, if $kT/\hbar\omega_c\gtrsim0.05$ the incompressible region is smeared out~\cite{siddiki2004,Oh97:13519}. Here, we present results only considering the quantized Hall plateau interval. In Fig.~\ref{fig:hall_bar}a, we show a situation where two incompressible strips reside along the edges and two encircling the inner contacts. The inset depicts cuts along the $x$ direction, where one sees that encircling ones are narrower than that of the \emph{edge} incompressible strips. Later we will argue that, once the incompressible regions become narrower than thermodynamical length scales they become evanescent and unable to decouple inner contacts. By decoupling, we mean that electron transport between two compressible regions is hindered by the incompressible region. Namely, scattering or tunneling is not possible across the stirp. At a higher $B$ field, we again observe that the 2DES comprises two edge incompressible strips parallel to the $y$- axis together with circular strips surrounding inner contacts (Fig.~\ref{fig:hall_bar}b), however the strips become wider. In this situation the bulk is compressible, in contrast, the inner contacts are decoupled from each other by the encircling strips, known as the Corbino effect~\cite{Dolgopolov92:12560}. At $\Omega_c=1.45$, the incompressible strips merge and the sample becomes approximately incompressible, as depicted in Fig.~\ref{fig:hall_bar}c. Increasing $B$ furthermore results in an electron distribution where the incompressible regions are disconnected around the contacts, as shown in Fig.~\ref{fig:hall_bar}d, however, are percolating along the edges. In this situation a dissipative transport between inner contacts is possible, which will be described by Ohm's law.

\subsection{Hall bar with disorder} The above situation is altered if one considers disorder. To analyze the disorder effects on the electron distribution, we embodied charged impurities to donor layer resulting in potential fluctuations. Following a recent investigation~\cite{sinem}, we considered sufficiently high number of impurities ($\sim 2000$). Note that, disorder is always included to our calculations that determines Landau level broadening and conductivities. Inclusion of the disorder, apparently enlarges the edge incompressible strips as shown in Fig.~\ref{hallbar_disorder}a. At a higher $B$ the two edge strips merge with the encircling strips as shown in Fig.~\ref{hallbar_disorder}b. By the inclusion of disorder, we observe that the bulk becomes incompressible for larger $B$ intervals, hence wider quantized Hall plateaus are expected in accordance with typical experiments~\cite{jose:epl}. A characteristic filling factor distribution is shown in Fig.~\ref{hallbar_disorder}c. Even grippingly, at a slightly higher $B$ field the bulk incompressible region starts to become disconnected around the inner contacts, allowing dissipative current paths between them.

\section{Transport between inner contacts} 
\begin{figure}[t!]
\includegraphics[width=1.\linewidth]{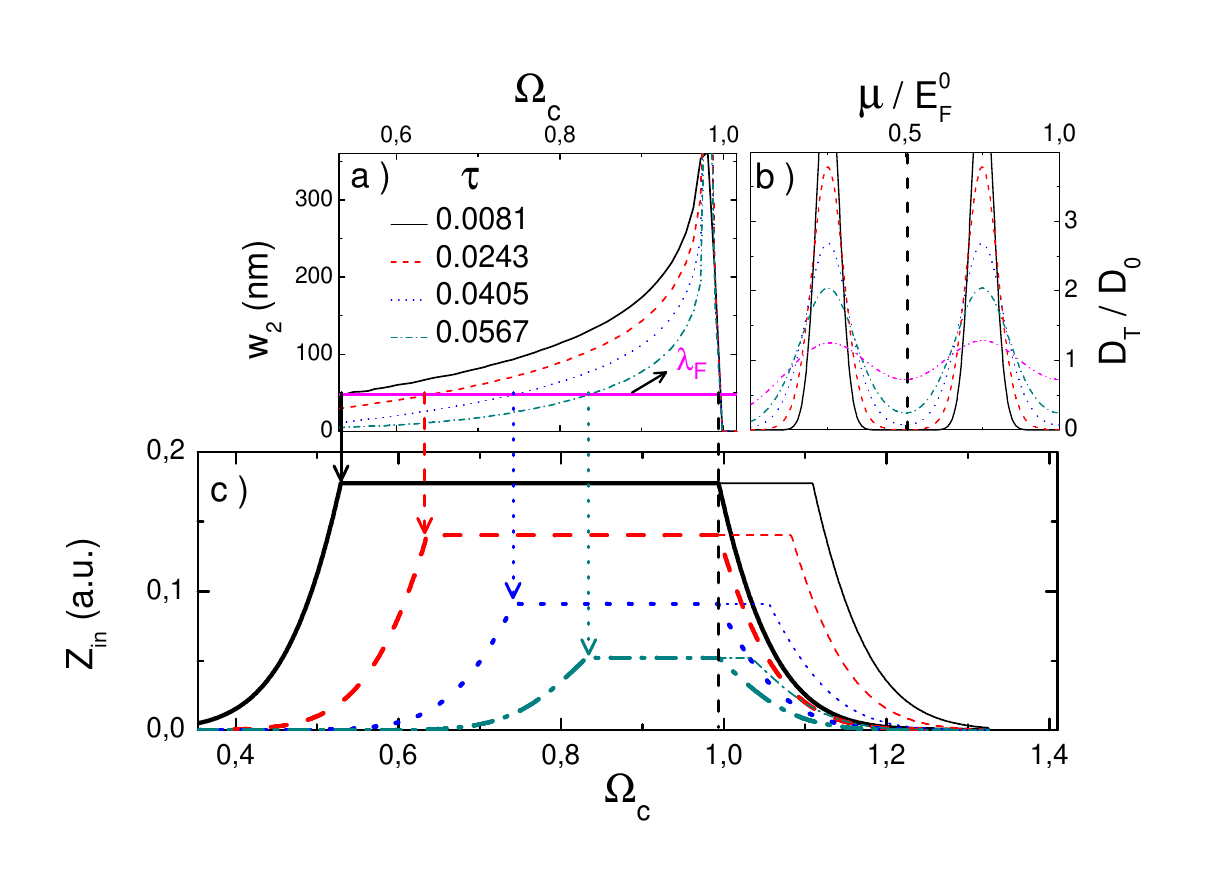}
\caption{\label{fig:capacitance} (a) Calculated widths of incompressible strips at different temperatures and (b) corresponding thermodynamical density of states in units of $D_0=2.8\times10^{10}$ meV$^{-1}$cm$^{-2}$. (c) Resulting impedance between inner contacts without (thick lines) and with potential fluctuations (thin lines).}
\end{figure}

As mentioned, incompressible bulk is a key concept in developing the analogy between topological insulators and conductance quantization. However, we have shown that it is possible to have a compressible metal-like bulk within the plateau interval due to interactions. Next, we investigate the transport between the inner contacts by calculating the impedance when a finite potential difference is applied between the inner contacts. In our model we assume that the transport can be well described by a local version of the Ohm's law~\cite{Guven03:115327} and the impedance is composed of resistive $R$ and capacitive $C$ terms as, $Z=\sqrt{R^2+\frac{a}{C^2}}$. Here, $a$ is a constant depending on the frequency of the imposed excitation, which we take to be of the order of unity. The calculation of the (quantum) capacitance is somewhat trivial, since it is only proportional to the thermodynamic density of states at $E_F$, namely $C=e^2D_T(E_F)$. If there exists an incompressible region (i.e. $D_T(E_F)=0$), which decouples inner contacts then the impedance reads to infinity for a pure Hall bar at $T=0$. However, for a realistic Hall bar with disorder, i.e. with localized states at $E_F$, and at finite temperatures capacitance also becomes finite. To decide whether if the inner contacts are decoupled or not by the encircling incompressible strips, our first task is to determine the widths of the incompressible regions depending on $B$, which we calculate self-consistently and show in Fig.~\ref{fig:capacitance}a. Once the strip widths are wider than the thermodynamical length scale, i.e. Fermi wavelength $\lambda_F$, one can assume that the inner contacts are decoupled. Hence, impedance is dominated by the capacitive term. Fig.~\ref{fig:capacitance}b depicts $D_T(E)$, obtained from Gaussian broadened Landau levels at different temperatures. The corresponding impedance is shown in Fig.~\ref{fig:capacitance}c, where we observe that $Z$ is constant ($\propto1/D_T(E_F)$) if incompressible regions decouple inner contacts. Otherwise, it is determined by the resistive term which we assume again a Gaussian level broadening to calculate $R(B)$, see Fig.~\ref{fig:capacitance}b. By increasing the temperature the incompressible strips become narrower, hence, the constant $Z$ interval shrinks while its amplitude decreases due to larger available states at $E_F$. We also depicted the disordered situation by thin lines in Fig.~\ref{fig:capacitance}c where the bulk incompressible region is extended in $B$ as shown previously. The result is wider constant $Z$ intervals, which decrease in amplitude and shrink in width similar to the pure Hall bar case by increasing the temperature.

\section{Conclusion}
Our numerical calculations show that the bulk of the electronic system is not insulating for all magnetic field strengths throughout the plateau interval and IQHE is insensitive to topological distortion in real space. This finding is in well agreement with existing local probe experiments, however, is in contrast with the single particle theories of the IQHE. The discrepancy is removed by the consideration of the disorder effects. In this work we quantitatively demonstrated that the edge and bulk pictures of the IQHE merge, which are both the limiting cases of the screening theory.

In light of our calculations we propose to measure Hall resistance of a 2DES that embodies two inner contacts in the bulk and simultaneously measure the potential difference between them. A constant high impedance $B$ subinterval is predicted within the plateau regime, in contrast to single particle theories for which one expects the constant $Z$ interval to be spread all over the plateau regime. We claim that the varying impedance evidences a compressible bulk. It is also expected that the amplitude of the constant $Z$ will decrease with increasing temperature. Different from the early experiments~\cite{Tsui:85:potdist} we suggest to consider a relatively narrow Hall bar ($d<20$ $\mu$m) to be defined on a high mobility ($> 3\times10^6$ cm$^{2}$/Vs) wafer.

\section{Acknowledgments}
We would like to thank J. Jain and K. von Klitzing for initiating the idea of inner contacts.
This work is supported by T\"UB\.ITAK under grant 112T264 and 211T148, IU-BAP:6970 and 22662.

\section*{References}
\bibliographystyle{unsrt}

\end{document}